\documentclass[conference]{IEEEtran}
\IEEEoverridecommandlockouts
\usepackage{cite}
\usepackage{amsmath,amssymb,amsfonts}
\usepackage{algorithmic}
\usepackage{graphicx}
\usepackage{textcomp}
\usepackage{xcolor}
\usepackage[misc]{ifsym}

\def\BibTeX{{\rm B\kern-.05em{\sc i\kern-.025em b}\kern-.08em
    T\kern-.1667em\lower.7ex\hbox{E}\kern-.125emX}}
\begin{document}

\title{EfficientASR: Speech Recognition Network Compression via Attention Redundancy and Chunk-Level FFN Optimization}

\author{\IEEEauthorblockN{Jianzong Wang$^{1\ddagger}$, Ziqi Liang$^{1,2\ddagger}$\thanks{$\ddagger$ Equal Contributions}, Xulong Zhang$^{1\textrm{\Letter}}$\thanks{$^\textrm{\Letter}$Corresponding author: Xulong Zhang (zhangxulong@ieee.org).}, Ning Cheng$^{1}$, Jing Xiao$^{1}$}
\IEEEauthorblockA{\textit{$^{1}$Ping An Technology (Shenzhen) Co., Ltd.}\\\textit{$^{2}$University of Science and Technology of China}}
}

% \author{\IEEEauthorblockN{Jianzong Wang$^{1\dagger}$, Ziqi Liang$^{1,2\dagger}$\thanks{$\dagger$ Equal Contributions}, Xulong Zhang$^{1\ast}$\thanks{$^\ast$Corresponding author: Xulong Zhang (zhangxulong@ieee.org).}, Ning Cheng$^{1}$, Jing Xiao$^{1}$}
% \IEEEauthorblockA{\textit{$^{1}$Ping An Technology (Shenzhen) Co., Ltd.}\\\textit{$^{2}$University of Science and Technology of China}}
% }

\maketitle

\begin{abstract}
In recent years, Transformer networks have shown remarkable performance in speech recognition tasks. However, their deployment poses challenges due to high computational and storage resource requirements. To address this issue, a lightweight model called EfficientASR is proposed in this paper, aiming to enhance the versatility of Transformer models. EfficientASR employs two primary modules: Shared Residual Multi-Head Attention (SRMHA) and Chunk-Level Feedforward Networks (CFFN). The SRMHA module effectively reduces redundant computations in the network, while the CFFN module captures spatial knowledge and reduces the number of parameters. The effectiveness of the EfficientASR model is validated on two public datasets, namely Aishell-1 and HKUST. Experimental results demonstrate a 36\% reduction in parameters compared to the baseline Transformer network, along with improvements of 0.3\% and 0.2\% in Character Error Rate (CER) on the Aishell-1 and HKUST datasets, respectively.
\end{abstract}

\begin{IEEEkeywords}
speech recognition, attention redundancy, feedforward network, lightweight
\end{IEEEkeywords}

\section{Introduction}

In recent years, Transformers \cite{transformer2017} have shown better performance than traditional sequence models \cite{RNN,LSTM} in the ASR domain, capturing long-range dependencies through attention mechanisms. Attention mechanisms play a crucial role in Transformers by establishing connections between different positions through identity mapping. However, attention computation in Transformers is computationally expensive and contains a significant amount of redundancy. On the other hand, feedforward networks capture high-level representations through high-dimensional feature mappings, but this also leads to an increase in network parameters. These issues result in high-performing Transformer models requiring substantial storage and computational resources, making it challenging to apply them on limited computing devices.

The importance of attention mechanisms in Transformer networks has been extensively studied. For instance, He et al. \cite{he2020realformer} introduced the residual attention mechanism to improve model performance by enabling interaction and fusion of attention across different layers. Research by \cite{usefulatten} revealed that as attention propagates from lower to higher layers, the attention distribution becomes more diagonalized, with higher-layer attention contributing less to model performance. Additionally, the study by \cite{randomhead} demonstrated that randomly removing some attention heads does not significantly affect model performance, suggesting redundancy in attention computations. Shim et al. \cite{shim2021understanding} found that lower-layer attention focuses more on semantic features of speech, while higher-layer attention is more concerned with language positioning. Moreover, the investigation by \cite{reuseatten} discovered that attention distributions between adjacent layers tend to be more similar, reducing network redundancy by sharing certain attention heads.
HybridFormer \cite{HyperConformer} proposed HyperMixer to replace multi-head self-attention, which can model local interaction and global interaction information respectively. Benefit from HyperMixer's linear time and memory complexity, significantly reducing computational and memory costs.

In the domain of model lightweighting, effective methods like quantization-aware training \cite{Q8Bert} and knowledge distillation \cite{Tinybert} \cite{DistilBERT} exist but can be complex to implement. In contrast, weight sharing \cite{lan2019albert} \cite{dehghani2018universal} offers a more straightforward model compression approach. 
However, when only the learnable parameters are shared, it doesn't reduce the forward propagation computation of the network. 
By analyzing the parameter and computational costs of the attention mechanism and feed-forward networks in Transformer, we introduce an innovative approach to address attention redundancy, sharing attention scores across layers through residual connections for interactive fusion between higher and lower layers.
At the same time, in order to reduce the computational redundancy introduced by cross-layer shared attention scores and enhance the ability of attention to extract local features, we apply the sliding window with deformability (SWD) method \cite{LFEformer} to shared attention scores, reducing redundancy in single-layer attention maps.
When lightweighting attention-based speech recognition models, we often focus on the point: Reduce attention redundancy and parameter count effectively.

To tackle these challenges, in this paper, we introduce an efficient ASR network, called EfficientASR, which 1) leverages sharing residual attention to reduce attention redundancy and 2) employs the Chunk-level FFN structure to reduce parameter count effectively. 
Specifically, the method shares attention scores among multiple layers and integrates old attention scores into new ones through residual connections, achieving interactive fusion of attention scores between higher and lower layers.
Chunk-level Feed-forward Neural Network (CFFN) divides the feed-forward networks into multiple chunks based on embedding dimensions, using smaller FFNs in each chunk, substantially reducing the number of learnable parameters without compromising model performance. 
Additionally, the paper applies the previously proposed  sliding window with deformability (SWD) \cite{LFEformer} method to the shared attention scores, addressing the slow change of diagonalization degree in higher layers after shared attention and further reducing redundancy in single-layer attention maps.

\section{Related Work}

\subsection{Automatic speech recognition}

With the application of transformer in the field of natural language processing (NLP) and its significant effect on modeling contextual information, transformer is also suitable for the field of speech \cite{Wang2019TransformerBasedAM,Watanabe2017HybridCA,ZhangLSTMKK20}.
Existing automatic speech recognition (ASR) works are mainly divided into three structures: CNN, RNN, and transformer. Previously, transformer and CNN were used as networks to achieve good results in ASR, but both have their limitations: transformer is not good at extracting fine-grained partial feature, while the CNN network can capture local features such as edge lines and shapes, but more convolutional layers are needed to capture global information.

In subsequent work on the ASR task, there are improvements in recognition efficiency by using depth-wise separable convolutions \cite{KrimanBGHKLLLZ20}. The model is composed of several interconnected modules, featuring residual links that facilitate the flow of information between them. Within each module, there is an implementation of one or multiple 1D temporal channel separable convolutional filters.
There is also a combination of squeeze and excitation modules to capture longer contextual semantic information \cite{han20_interspeech}.
Compared with CNN, the significant benefit of transformer is that it can learn global dependencies based on self-attention, which is key for speech processing tasks.
Gulati et al.~\cite{conformer} proposed the Conformer, since the Transformer model is good at capturing global interactions based on content and CNN effectively utilizes local features, Conformer integrates the capabilities of both the Transformer and CNN to effectively capture and process both local and global interdependencies present within the audio sequence.

However, in the previous Conformer structure, the feature representations derived from consecutive speech frames contained considerable repetition, resulting in unnecessary computational consumption. In this regard, Squeezeformer \cite{Squeezeformer} introduces a sequential U-Net structure. The down-sampling layer halves the sampling rate of the input speech signal. At the end of the network, a lightweight up-sampling layer restores the speech signal from a low sampling rate to the original high sampling rate, which ensures that model training is stable. Branchformer \cite{Branchformer} adopts a parallel dual-branch structure. 
A particular branch within the model leverages a multi-headed self-attention system to capture expansive characteristics across the input sequence, and the other branch introduces the MLP with convolutional gating structure, which is intended to capture local features in the audio sequence. \cite{E-Branchformer} proposed E-Branchformer, an advancement over the Branchformer, which applys efficient model fusion methods and stacking additional point-adding modules. As a result, E-Branchformer established a new benchmark for WER, notably without leveraging any external training data.
Zipformer \cite{Zipformer} has similar ideas to previous work on downsampling the temporal dimension. However, compared to the fixed downsampling ratio in the Squeezeformer \cite{Squeezeformer}, the Squeezeformer works with different downsampling ratios in different encoder stacks, and uses a larger downsampling ratio in the intermediate encoder stack.
% Inspired by Conformer \cite{conformer}, EfficientConformer \cite{Efficient_Conformer} introduces progressive downsampling into the Conformer encoder, and proposes a grouped attention mechanism, which reduces the computational complexity of the self-attention module from $O(n^{2}d)$ to $O(n^{2}d/g)$, n is the time dimension, d is the hidden layer dimension, and $g$ is the group size.

\subsection{Efficient attention mechanism}
The self-attention mechanism is a critical component in many ASR models. It allows the model to dynamically focus on different parts of the input sequence, which is essential for understanding the context and structure of the speech data. However, the computational complexity of this mechanism is a significant concern due to its quadratic scaling with respect to the length of the input sequence. 
There are many ways to improve the efficiency of the transformer. Sparse attention mechanisms~\cite{TayBMJZZ21, ZaheerGDAAOPRWY20,Longformer,ZaheerGDAAOPRWY20,ZhangTWCLX23,Efficient_Conformer,ZhangPLG23,dhawan-etal-2023-unified} offer a promising approach to address the challenges associated with the complexity of self-attention.
% Some methods \cite{TayBMJZZ21, ZaheerGDAAOPRWY20,Longformer,ZaheerGDAAOPRWY20,zhang2022linguistic} use sparse attention mechanisms to reduce the complexity of self-attention. 
% \cite{TayBMJZZ21} studies the real contribution of the dot product-based self-attention mechanism to the performance of the Transformer, and proposes Synthesizer that can learn synthetic attention weights without token-to-token interaction. 
% It achieves highly competitive performance compared to ordinary Transformer models across a range of tasks, including machine translation, language modeling, and text generation.
Through the self-attention process, the mechanism transcends limitations inherent in RNNs' sequential framework, allowing individual tokens in a given sequence to engage with the entire set of tokens without dependence on their order. 

However, numerous applications limit context to a very short sequence length due to current hardware and model size constraints. These constraints typically restrict input sequences to approximately 512 tokens in length, significantly diminishing direct applicability to tasks requiring broader context. To tackle this issue, \cite{ZaheerGDAAOPRWY20} propose a sparse attention mechanism capable of processing sequences up to 8 times longer than previously achievable with comparable hardware. This advancement significantly enhances the performance of various Natural Language Processing (NLP) tasks, including question answering and summarization, by effectively managing longer in-context information.
Longformer \cite{Longformer} uses sliding window attention to focus on local in-context information, and uses global attention on some preselected input positions to capture global in-context information. 
To mitigate potential declines in performance, \cite{ZaheerGDAAOPRWY20} focuses on randomly selected pairs of tokens, but it might necessitate utilizing a substantial quantity of tokens to reduce performance degradation on longer sequences, which often limits computational speed-up.

% \begin{figure*}[htbp]
%     \centering
%     \begin{minipage}[t]{0.49\linewidth}
%     \centering
%     \includegraphics[width=\textwidth]{figures/transformer_page-0001.jpg}
%     \centerline{(a)}
%     \end{minipage}
%     \begin{minipage}[t]{0.50\linewidth}
%     \centering
%     \includegraphics[width=\textwidth]{figures/chunk_ffn_page-0001.jpg}
%     \centerline{(b)}
%     \end{minipage}
%     \caption{ (a) is a traditional Transformer structure.
%     (b) is an EfficientASR structure, where the dashed line portion is only used in shared attention mode. The small circles represent the embedding dimension of the input features, and the small squares represent learnable parameters. SWD represents the  sliding window with deformability}
%     \label{overall}
%     \vspace{-10pt}
% \end{figure*}

\begin{figure*}[h]
    \centering
    % \vspace{-0.2cm}
    \includegraphics[width=18.2cm]{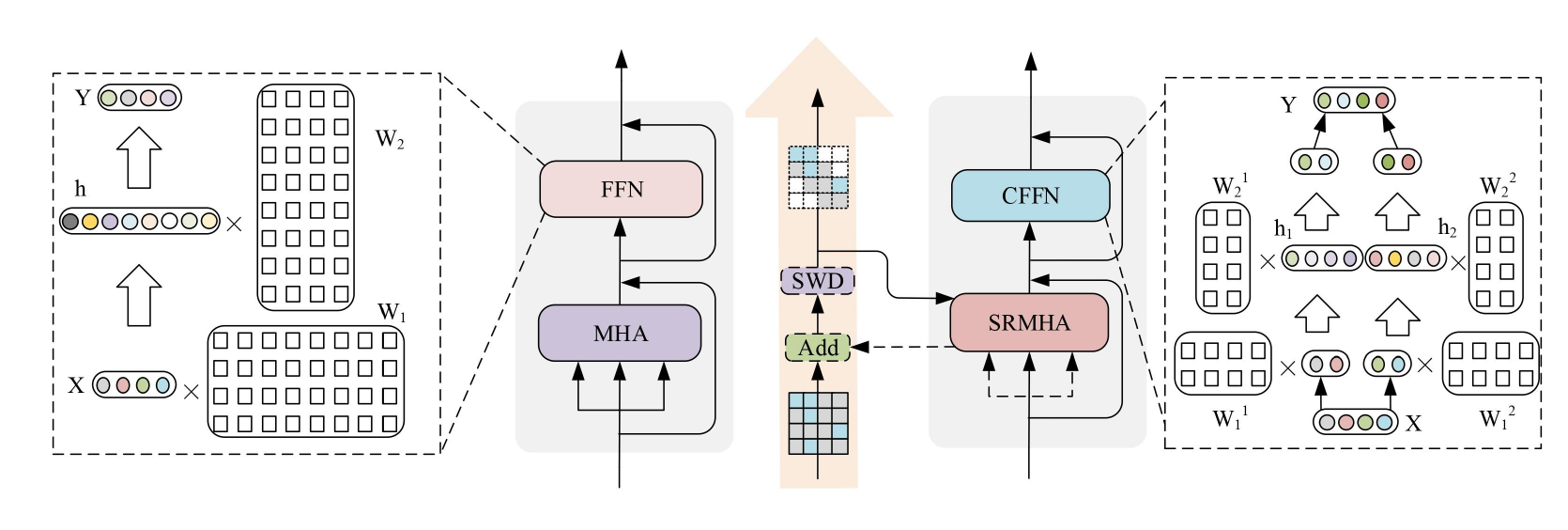}
    \vspace{-0.8cm}
    \caption{(a) is a traditional Transformer structure. (b) is an EfficientASR structure, where the dashed line portion is only used in shared attention mode. The small circles represent the embedding dimension of the input features, and the small squares represent learnable parameters. SWD represents the  sliding window with deformability}
    % \vspace{-0.3cm}
    \label{overall}
\end{figure*}

% In real-world use cases, access to extensive datasets that are manually annotated, along with ample computational capabilities, facilitates the training of sophisticated deep neural networks that boast numerous parameters and robust functionalities. Such networks are capable of attaining remarkably low Word Error Rates (WER) when evaluated on scholarly benchmark tests. 
However, even if these new methods surpass the state-of-the-art, their use in the real world often comes with significant resource costs. Hence, there is a crucial need to investigate efficient and lightweight model networks that hold increased commercial value.

% \vspace{0.2cm}
\section{Proposed Method}

\subsection{Review Transformer}

The encoder-decoder Transformer model has achieved excellent performance in ASR, a success largely attributable to its integration of an attention mechanism with a feed-forward neural network. The specific calculation process of the attention mechanism is as follows: 
\begin{equation}\label{W}
    Attn= Softmax(\frac{{Q}{K^{T}}}{\sqrt{d_k}})V,
\end{equation}

The number of learnable parameters for each attention module is $4d_{model}^{2}+4d_{model}$. In addition, when the input $X\in \mathbb{R}^{B\times{T}\times d_{model}}$ passes through the attention module, the number of floating point calculations is $4T^{2}Bd_{model} +8BTd_{model}^{2}$.

For the feed-forward network, it consists of two linear layers, where high-dimensional feature mapping is usually applied:
\begin{equation}
    FFN(X) = ReLU(XW_1+b_1)W_2+b_2,
\end{equation}
% Where $ReLU$ serves as the activator, the matrices $W_1\in \mathbb{R}^{d_{model}\times d_{ff}}$, $W_2\in \mathbb{R}^{d_{ff}\times d_{model}}$ are trainable parameter matrices, and $d_{ff}$ is the hidden dimension of FFN. 
Where $ReLU$ serves as the activator, the matrices $W_1$, $W_2$ are belonging to the space $\mathbb{R}^{d_{model}\times d_{ff}}$ and $\mathbb{R}^{d_{ff}\times d_{model}}$, and $d_{ff}$ is the hidden dimension of FFN. 
$b_1\in \mathbb{R}^{d_{ff}}$ and $b_2\in \mathbb{R}^{d_{model}}$ are bias vectors. %Therefore, the number of parameters in each FFN module is $2d_{model}d_{ff}+d_{model}+d_{ff}$. When the input $X\in \mathbb{R}^{B\times{T}\times d_{model}}$ passes through an FFN module, it will generate computational complexity $4BTd_{ff}d_{model}$. 
It is worth noting that $d_{ff}$ is often several times greater than $d_{model}$, for example, in the ESPnet \cite{watanabe2018espnet} toolkit, the default value of $d_{ff}$ is $8d_{model}$. Therefore, the parameter count of the FFN is $16d_{model}^2+9d_{model}$, and the number of floating point calculations is $32BTd_{model}^2$.

Based on the above analysis, it can be found that when $d_{ff}= 8d_{model}$, the number of learnable parameters in FFN is much larger than that of MHA, which is caused by the large weight matrices in FFN due to high-dimensional feature mapping. 
% Although the number of learnable parameters in MHA is smaller than FFN, we found that when the length $T$ of the input features becomes larger, the computational complexity of MHA will be much higher than that of FFN. 
Even though MHA possesses a lesser count of adjustable parameters compared to FFN, our research indicates that as the size of the input feature set, represented by length $T$, increases the computational demand for MHA escalates significantly, surpassing that reguired by FFN.
This is because the computational complexity of ${Q}{K^{T}}$ and $SV$ in the calculation process is proportional to the square of $T$. Therefore, this also inspires us to improve both MHA and FFN to reduce the overall computational complexity and number of learnable parameters of the model.

\subsection{EfficientASR}

Following the preceding discussion, we introduce an innovative lightweight network structure called EfficientASR. Illustrated on the right side of Fig~\ref{overall}, each encoder layer is comprised of an SRMHA component, an SWD component, and a CFFN component, while the decoder layers also include a cross multi-head attention module. 
Residual links and layer-wise normalization are integrated within each module, as referenced in \cite{He2016DeepRL, Ba2016LayerN}. EfficientASR reduces the computational complexity and number of learnable parameters by reducing redundant information in the attention mechanism.

Previous studies have shown that there is a significant amount of redundancy in the repeated calculation of attention scores in Transformer networks \cite{randomhead} \cite{reuseatten}. However, simply sharing attention heads to other layers is not the optimal choice. 
Inspired by the concept of \cite{usefulatten} as the depth of the network increases, the diagonalization of attention becomes more prominent. Sharing attention heads directly to other layers slows down the formation of this diagonalization. 
In place of the conventional Transformer's multi-head self-attention component, we have introduced a substitute module with a shared residual multi-head attention (SRMHA) module, while retaining the cross self-attention in the decoder. The SRMHA module has two modes of operation across different sub-layers: shared attention and updated attention, as depicted on the right side of Figure~\ref{overall}. For any layer of SRMHA, its expression is:
\begin{equation}\label{qkv_s}
    S = W_{q}Z_{q}\cdot(W_{k}Z_{k})^{T}
    % \vspace{0.3cm}
\end{equation}
\begin{equation}\label{qkv_v}
    V = W_{v}Z_{v}
    % \vspace{0.3cm}
\end{equation}
\begin{equation}\label{QKV}
    SRMHA=Softmax(S)V.
    % \vspace{0.3cm}
\end{equation}
Here, $S$ represents attention scores, and $V$ represents the linear mapping of inputs.

\subsection{Updated Attention Mode}
% {\bf Updated Attention Mode:} 
In the speech recognition task, neighboring tokens have stronger correlations. Therefore, the SWD module restricts the interaction range between tokens by only retaining the data within the w-range of the attention matrix diagonal. Under the updated attention mode, the attention matrix $S$ is updated as follows:
\begin{equation}\label{QKV}
S=SWD(S', w),
% \vspace{0.3cm}
\end{equation}
\begin{equation}\label{QKV}
S'={\frac{{Q}{K^{T}}}{\sqrt{d_k}}}+S'',
% \vspace{0.3cm}
\end{equation}
here, $S'$ represents the updated attention matrix, $S''$ represents the previous updated attention matrix, and $d_{k}$ denotes the size of either the word vector dimensions or the dimensions within the hidden layer. By using residual connections for attention, similar to \cite{he2020realformer}, high-level and low-level feature distributions can be fused. Specifically, by connecting the attention matrix $\frac{{Q}{K^{T}}}{\sqrt{d_k}}$ generated from the input features of the current layer with the old attention matrix $S''$ through residual connections, we obtain a new attention matrix $S'$. The SWD module and attention matrix residual connections are only used in sub-layers under the updated attention mode.

\subsection{Shared Attention Mode}
% {\bf Shared Attention Mode:} 
To reduce the redundant calculation of attention, we reduce the need to calculate attention matrices for the same input features by sharing attention. Under the shared attention mode, the attention matrix $S$ is equal to the old attention matrix $S''$, and the SRMHA module does not need to generate the $Q$ and $K$ matrices, thereby reducing the computational burden of two linear layers and the calculation of $QK^{T}$. In this case, $S = S''$ , and the SRMHA module does not need to generate the $Q$ and $K$ matrices, reducing the computational cost of two linear layers and the calculation of $QK^{T}$. In this case, the SRMHA module has $2d_{model}^{2}+2d_{model}$ learnable parameters and requires $2T^{2}Bd_{model}+4BTd_{model}^{2}$ floating-point operations in terms of computation.
Compared with the standard self-attention module in Transformers, the learnable parameters and floating-point operations of the SRMHA module are reduced by half.

\subsection{Chunk-level Feed-Forward Network}
% {\bf Chunk-level Feed-Forward Network:} 
The mapping of high-dimensional features dominates most of the learnable parameters in Transformer, but it has been rarely studied in speech recognition. Based on the inspiration from \cite{kvmemory}, we treat the feed-forward network as a key-value memory and redefine its expression as:
\begin{equation}\label{chunk}
    FFN=f(XK^{T})V.
    % \vspace{0.3cm}
\end{equation}
Here, $K$ and $V$ correspond to the learnable matrices $W_1$ and $W_2$, respectively, and $f$ is utilized to signify the ReLU. The key of each layer is used to capture patterns in the input sequence, while the value generates the distribution of the next token based on the captured patterns. In order for the feed-forward network to serve as a key-value memory and learn different feature representations, covering a broader range of semantic relationships, we propose a chunk-level feed-forward network (CFFN). As shown in Fig. \ref{overall}, CFFN can project and map the input sequence differently, reducing the burden of high-dimensional computations and improving scalability. Specifically, the expression of CFFN is as follows:
\begin{equation}\label{QKV}
    CFFN=Concat(chunk_1,...,chunk_n),
    % \vspace{0.3cm}
\end{equation}
\begin{equation}\label{QKV}
    chunk_i=FFN(X_i),
    % \vspace{0.3cm}
\end{equation}
\begin{equation}\label{QKV}
    X_1,...,X_n=Split(X),
    % \vspace{0.3cm}
\end{equation}
Here, $n$ represents the number of chunks, and each chunk represents a key-value memory. When $d_{ff}= 8d_{model}$, the amount of parameters in each chunk is $16({d_{model}}/{n})^2+9d_{model}/n$. By introducing the chunk-level feed-forward network (CFFN), the model can better utilize the feed-forward network as a key-value memory to learn different feature representations, improve the performance of the speech recognition model, and reduce the burden of high-dimensional computations.

\subsection{The global learning of EfficientASR}
We selected 2 loss functions for global learning and the fitting of EfficientASR. We utilize cross-entropy loss $\mathcal{L}_{CE}$ to calculate the probability distribution difference between the EfficientASR output and the target sequence. A CTC loss $\mathcal{L}_{CTC}$ \cite{CTC} is selected to improve the recognition accuracy of acoustic boundaries and achieve the alignment of acoustic tokens and semantic tokens.

\begin{equation}\label{loss}
    \mathcal{L} = \alpha_{1}\mathcal{L}_{CE} + \alpha_{2}\mathcal{L}_{CTC}
    % \vspace{0.3cm}
\end{equation}
Notably, $\alpha_{1}$ and $\alpha_{2}$ represent weight hyperparameters that adjust the relative importance of these two loss functions. Once $\mathcal{L}$ is obtained, we compute the gradients of all parameters and update our model using backward propagation.

\section{Experiment}

\subsection{Experimental Setup}\label{sec4_1}
{\bf Dataset and Preprocessing:} To assess the performance of the EfficientASR, we have chosen two prominent datasets: Aishell-1 \cite{Aishell1} and HKUST \cite{liu2006hkust}. The preprocessing stage of the audio input involves a series of transformations, such as segmentation into frames, application of a window function, computation via the fast Fourier transform (FFT), and subsequent processing with the discrete cosine transform (DCT). In this process, each window spans 25 ms, with a stagger of 10 ms between successive windows. The resultant feature set consists of 80 dimensions, represented by log Mel-filterbank coefficients.
In the context of speech enhancement \cite{specaugent}, the time mask window is 30 and the frequency mask window is 40. In the convolutional downsampling, the size of the convolutional kernel is 31.

% {\bf Dataset and Preprocessing:} Aishell-1 \cite{Aishell1} and HKUST \cite{liu2006hkust} are used to validate the effectiveness of the proposed the EfficientASR network. The input audio signal undergoes feature transformation in the frontend, including framing, windowing (25ms window size and 10ms window offset), FFT, and DCT, resulting in 80-dimensional log Mel-filterbank coefficients. For speech enhancement \cite{specaugent}, we use a time mask window of 30 and a frequency mask window of 40. In convolutional downsampling, a 31-sized convolutional kernel is employed.

{\bf Hyperparameter:} The ESPNet toolkit~\cite{watanabe2018espnet} serves as the platform for executing all our experiments. Specifically, we have implemented the label smoothing technique and a dropout rate of $p$=0.1 to mitigate the risk of overfitting in our model. For optimization, we have chosen the Adam algorithm~\cite{2014Adam}, with a learning rate set to $0.002$.
Supplementary settings encompass a model dimension of $d_{model}$=256, a feed-forward layer size of $d_{ff}$=2048, with window dimensions denoted by $w$=6 and $h$=4, followed by the number of attention heads $12$ and the number of encoder layers $6$. 
In the attention mechanism, we have set $d_{q}$, $d_{k}$, and $d_{v}$ dimensions equal to $d_{model}/h$=64. Each training batch commences with a size of 64, and the beam search strategy employs a beam width of $10$. Our language model (LM), which operates on the Transformer, comprises $16$ lavers and is trained over $15$ epochs.
During training, the weight for the CTC is $0.3$, and during inference, the weight for CTC is $0.6$.

\subsection{Evalution systems}
To objectively compare the automatic speech recognition (ASR) performance of the acquired representations across various noisy environments, we developed four replication-synthesis frameworks utilizing both Aishell-1 \cite{Aishell1} and HKUST \cite{liu2006hkust} corpora:
\begin{itemize}
\item Transformer w/o LM \cite{transformer2017}: Transformer-based ASR model combined with pre-trained language model.
\item Transformer\_LM \cite{transformer2017}: Transformer-based ASR model combined without pre-trained language model
\item Conformer \cite{gulati2020conformer}: Combining CNN and transformer to model both local and global interdependencies within audio sequences.
% \item EfficientASR: Our proposed improved transformer-based ASR model based on shared residual multi-head attention (SRMHA) and chunk-level feedforward networks (CFFN).
\item Conformer\_EfficientASR: Replace multi-head attention (MHA) in Conformer with SRMHA and feedforward (FFN) network with CFFN.
\end{itemize}

\subsection{Results}\label{sec4_2}

The experimental results conducted on the Aishell-1 dataset are shown in Table~\ref{table1}. The EfficientASR model has only 19.33M parameters, which is a 36\% reduction compared to the Transformer model. The EfficientASR model realized a 0.1\% reduction in Character Error Rate (CER) for the development (dev) set and a 0.3\% decrease for the test sets. To validate the model's scalability, SRMHA and CFFN were applied to the Conformer model, resulting in the Conformer\_EfficientASR model. Experimental results indicate that the Conformer\_EfficientASR model reduces the parameter count by 38\% compared to the Conformer model while maintaining a test set CER of 4.9\%. Therefore, the proposed EfficientASR model significantly minimizes the quantity of parameters that require learning, as well as the computational expense of the network, without notably degrading the performance of the model.

\begin{table}[!t]
% \vspace{-0.5cm}
	\centering
		\caption{Comparison of CER results between EfficientASR and Transformer baseline on Aishell-1.} 
  		\vspace{-0.2cm}
	\begin{tabular}{lcc}
		\hline
		Model & dev/test (\%) & Parameter (M) \\
		\hline
           % Realformer \cite{he2020realformer} & 5.2 &  5.4 & 30.36 \\
           % Attention Map Reuse\cite{reusemap} & 5.0 &  5.2 & 30.36 \\
           % Reuse Transformer \cite{reuseatten}& 7.8 &  4.7 & 30.36 \\
           Transformer w/o LM \cite{transformer2017} & 5.5/5.9 & 30.35   \\
           Transformer\_LM\cite{transformer2017} & 5.2/5.6 & 30.35  \\
           Conformer \cite{gulati2020conformer} & 4.5/4.9 & 45.40  \\
        \hline
        EfficientASR w/o LM & 5.4/\textbf{5.6} & 19.33 (\textbf{36\%} $\downarrow$) \\
        EfficientASR\_LM & 5.1/\textbf{5.3}  & 19.33 (\textbf{36\%} $\downarrow$) \\
        Conformer\_EfficientASR & 4.6/4.9 & 28.10 (\textbf{38\%} $\downarrow$) \\
        \hline\\
	\end{tabular}
	\label{table1}
	\vspace{-0.5cm}
\end{table} 

\begin{table}[!t]
	\centering
		\caption{Comparison of CER results between EfficientASR  and Transformer baseline in HKUST dataset.} 
		\vspace{-0.2cm}
	\begin{tabular}{lcc}
		\hline
		Model & dev (\%) & Parameter (M) \\
		\hline
           % Realformer \cite{he2020realformer} & 21.8 &  29.85  \\
           % Attention Map Reuse\cite{reusemap} & 8.9 &  5.9  \\
           % Reuse Transformer \cite{reuseatten}& 7.8 &  4.7 \\
           Transformer w/o LM \cite{transformer2017} & 21.7 & 29.85   \\
           Transformer\_LM\cite{transformer2017} & 21.5 &  29.85  \\
           Conformer \cite{gulati2020conformer} & 19.8 & 44.96 \\
        \hline
        EfficientASR w/o LM & 22.0 & 18.86 (\textbf{36\%} $\downarrow$) \\
        EfficientASR\_LM & \textbf{21.3} & 18.86 (\textbf{36\%} $\downarrow$) \\
        Conformer\_EfficientASR & 20.1 & 29.23 (\textbf{38\%} $\downarrow$) \\
        \hline\\
	\end{tabular}
	\label{table2}
	\vspace{-0.6cm}
\end{table} 

Similarly, we further validated the proposed EfficientASR model on the HKUST dataset, and the experimental results are presented in Table~\ref{table2}. In contrast to the Transformer model, the EfficientASR model reduces the parameter count by 36\%. 
Employing a Language Model (LM) on the dev set led to a 0.2\% enhancement in CER performance.
In contrast to the Conformer model, the Conformer\_EfficientASR achieves a reduction of 38\% in the count of parameters, but lowers the CER by 0.3\%. These experiments on different datasets demonstrate that the EfficientASR model maintains strong competitiveness.

\begin{table}[h]
\vspace{-0.3cm}
	\centering
		\caption{Comparison of SRMHA and Other Attentional Redundancy Reduction Methods on HKUST Dataset} 
		\vspace{-0.2cm}
	\begin{tabular}{lcc}
		\hline
		Model & dev (\%) & Parameter (M) \\
		
		\hline
           Transformer \cite{transformer2017} & 21.5 &  29.85  \\
           Realformer \cite{he2020realformer} & 21.5 &  29.85  \\
           SPAH \cite{reuseatten}& 21.3&  29.06  \\
           SEAM\cite{reusemap} & 21.4 &  28.40  \\
           RSEAM  & 21.3 & 28.40   \\
        \hline
        SRMHA & \textbf{21.0} & 28.40 \\
        \hline\\
	\end{tabular}
	\label{table3}
	\vspace{-0.6cm}
\end{table} 

\subsection{Ablation Studies}\label{sec4_3}

{\bf Method comparison:} We compared the sharing of partial attention heads (SPAH) \cite{reuseatten} with sharing the entire attention map (SEAM) \cite{reusemap} for reducing attention redundancy. We also compared the addition of residual attention based on sharing the entire attention map (RSEAM), as well as the use of the SWD method in SRMHA.

The results of the experiments are detailed in Table \ref{table3}. The performance of residual attention is comparable to the baseline Transformer. 
We evaluated the model’s CER on the development set and the parameter size of different models.
SPAH improved the CER by 0.2\% while reducing only 0.79M parameters. SEAM improved the CER by only 0.1\% but reduced 1.45M parameters and reduced more attention computations. Based on these experimental results, combining residual attention with sharing the entire attention map can simultaneously reduce attention redundancy and improve network performance. Furthermore, when applying the SWD method to the network, the network can reduce even more redundant computations and exhibit diagonalized features at higher layers. 
Therefore, compared to previous methods, SRMHA has more advantages as it not only eliminates more attention redundancy but also achieves better performance in model performance.

\begin{table}[t!]
% \vspace{-0.3cm}
	\centering
		\caption{Use SRMHA on the encoder or decoder on the HKUST dataset.} 
		\vspace{-0.1cm}
	\begin{tabular}{lccc}
		\hline
		Encoder & Decoder & dev(\%) & Parameter (M) \\
		
		\hline
           $SR_2$  & $\times$ &  22.1 & 29.06  \\
           $SR_3$  & $\times$ &  21.1 & 28.79 \\
           $SR_4$ & $\times$ &  21.1  & 28.66\\
           $SR_6$  & $\times$ & 21.3  & 28.53 \\
           $SR_{12}$ & $\times$& 21.8 & 28.40 \\
           $\times$ & $SR_2$ &  21.2  & 29.45 \\
           $\times$ & $SR_3$ &  21.4  & 29.32\\
           $\times$ & $SR_6$ &  21.2  & 29.19 \\
           $SR_3$  & $SR_2$ &  21.0 & 28.40 \\
        \hline
	\end{tabular}
	\label{table4}
	\vspace{-0.5cm}
\end{table} 

In order to examine the effects of minimizing attention mechanism calculations on the model's performance, we integrated the SRMHA technique into both the encoding and decoding components of the Transformer architecture. The findings from our experiments are delineated in Table \ref{table4}.

{\bf Attention Redundancy Analysis:} To investigate the effects of reducing attention computations on the model, we integrated the SRMHA method into both the encoding and decoding components of the Transformer. The findings from our experiments are delineated in Table~\ref{table4}. Here, $SR\_i$ denotes updating attention computations every $i$ sub-layers, where a larger value of $i$ leads to a reduction in parameters and attention computations. $\times$ represents the traditional self-attention method.

In the encoder, using SRMHA with updates every 3 or 4 layers yielded optimal results, and updating every 6 layers did not significantly impact model performance. Similarly, in the decoder, using SRMHA with values of $i=2$, $i=4$, and $i=6$ all benefited network performance. This is because by employing SRMHA in the decoder instead of multi-head attention (MHA), the contextual distribution of labels can be repeatedly applied at different layers, thereby facilitating network learning.

Ultimately, in EfficientASR, we selected a configuration of $i=3$ in the encoder and $i=2$ in the decoder. Through this set of experiments, we found that SRMHA did not significantly reduce the majority of network parameters. However, the method analysis revealed that it could effectively reduce a substantial amount of repetitive attention computations.

\begin{table}[h]
\vspace{-0.2cm}
	\centering
		\caption{Compare the impact of different chunk-level FFNs.} 
		\vspace{-0.1cm}
	\begin{tabular}{lccc}
		\hline
		Model & Dataset & CER (\%) & Parameter (M) \\
		\hline
        Transformer \cite{transformer2017}&Aishell-1 & 5.2/5.6 & 30.35   \\
        1/2FFN & Aishell-1 &  \textbf{5.2/5.4} & \textbf{20.91} (\textbf{31\%} $\downarrow$) \\
        1/4FFN & Aishell-1 &  \textbf{5.3/5.6} & \textbf{16.20} (\textbf{47\%} $\downarrow$) \\
        \hline        Transformer\cite{transformer2017}& HKUST & 21.5 &  29.85  \\
        1/2FFN & HKUST &  \textbf{21.6} & \textbf{20.44} (\textbf{32\%} $\downarrow$) \\
        1/4FFN & HKUST &  22.4 & \textbf{15.74} (\textbf{47\%} $\downarrow$) \\
        \hline\\
	\end{tabular}
	\label{table5}
	\vspace{-0.3cm}
\end{table}

{\bf Chunk-level Feedforward Networks:} We partitioned the embedding dimension of input features into multiple chunks and analyzed the effects of varying chunk sizes on the efficiency of performance. The results are detailed in Table~\ref{table5}, where 1/2 FFN and 1/4 FFN represent splitting the embedding dimension into two and four chunks, respectively.

On the Aishell-1 corpora, the results showed that using 1/2FFN reduced the network parameters by 31\% and improved the CER by 0.2\% on the test sets. Further employing 1/4FFN reduced 47\% of parameters while achieving the same CER as the baseline.
However, on the HKUST dataset, using 1/2FFN reduced parameters by 32\%, but performance deteriorated with a 1\% increase in CER. On the other hand, using 1/4FFN reduced 47\% of parameters, but performance experienced a significant decline.
% The experimental results on two datasets (Aishell-1, HKUST) showed that using 1/2 FFN reduced the network parameters by (31\%, 32\%) and improved the Character Error Rate (CER) by (0.2\%, 1\%) on the test dataset. Further employing 1/4 FFN reduced (47\%, 47\%) of parameters while achieving the CER by (0.1\%, 0.9\%) higher than the baseline. 
Based on the experimental results, we observed significant fluctuations in the effectiveness of this method across different datasets. Nevertheless, selecting 1/2 FFN allows for a significant reduction in network parameters while maintaining stable performance.

\subsection{Memory usage of long sequences}
\begin{figure}[h]
    \centering
    \vspace{-0.2cm}
    \includegraphics[width=7.5cm]{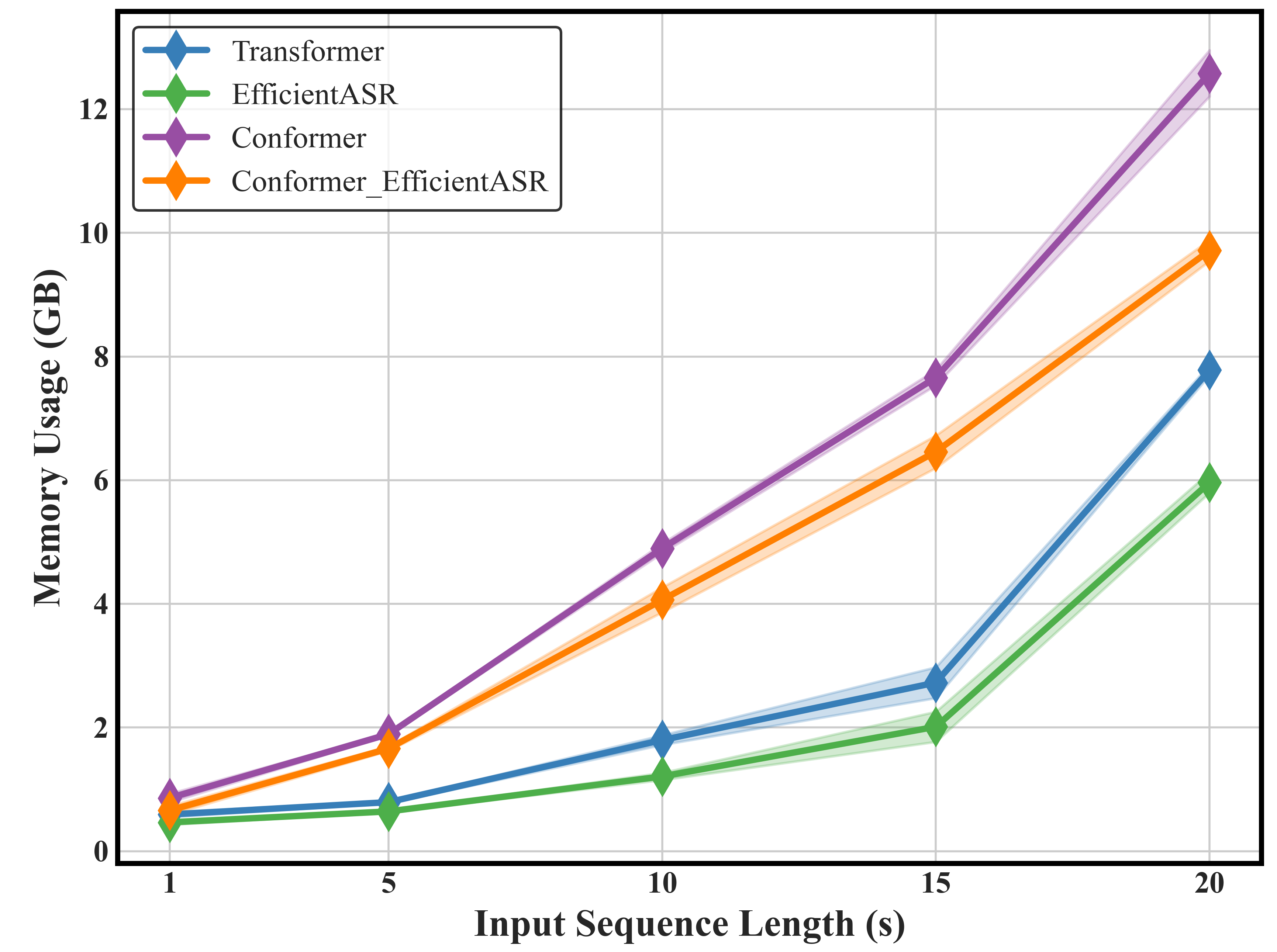}
    \vspace{-0.2cm}
    \caption{Different models' memory usage for processing long sequences on HKUST dataset.}
    % \vspace{-0.3cm}
    \label{memory}
\end{figure}

As shown in Fig.\ref{memory}, we show the impact of using SRMHA and CFFN on memory usage at different input sequence lengths. Experimental results show that as the sequence length continues to increase, the architecture using EfficientASR can effectively reduce the overall memory consumption (including model loading and data reading).

For audio sequences longer than 10 seconds, the utilization of SRMHA and CFFN in the encoder and decoder layers of ASR model leads to a reduction in overall memory usage for the same sequence length. Regardless of whether the base model is a Transformer-based or Conformer-based model, our method demonstrates the ability to decrease memory consumption when processing long sequences.

\section{Conclution}

The proposed EfficientASR, 1) reduces redundant computations, enhances attention diagonalization and fusion with sliding windows and residuals in the network; 2) captures spatial knowledge and reduces parameter size by dividing the input features into multiple chunks and using smaller feed-forward networks in each chunk.

\section{Acknowledgement}

Supported by the Key Research and Development Program of Guangdong Province (grant No. 2021B0101400003) and Corresponding author is Xulong Zhang (zhangxulong@ieee.org).

\bibliographystyle{IEEEtran}
\bibliography{reference}

\end{document}